\documentclass{egpubl}
\usepackage{egsr2023}
 
\SpecialIssuePaper         

\CGFStandardLicense

\usepackage[T1]{fontenc}
\usepackage{dfadobe}  

\usepackage{cite}  
\BibtexOrBiblatex
\electronicVersion
\PrintedOrElectronic
\ifpdf \usepackage[pdftex]{graphicx} \pdfcompresslevel=9
\else \usepackage[dvips]{graphicx} \fi

\usepackage{egweblnk} 

\usepackage{braket}
\usepackage{graphicx}
\usepackage{amsmath}
\usepackage{bm}
\usepackage{algorithm}
\usepackage{algorithmicx}
\usepackage{algpseudocode}
\usepackage{multirow}
\usepackage{booktabs}
\usepackage{pifont}

\newcommand{\cmark}{\ding{51}}%
\newcommand{\xmark}{\ding{55}}%

\title[Neural Free-Viewpoint Relighting for Glossy Indirect Illumination]%
      {Neural Free-Viewpoint Relighting for Glossy Indirect Illumination}

\author[N. Raghavan$^*$, Y. Xiao$^*$, K.-E. Lin, T. Sun, S. Bi, Z. Xu, T.-M. Li and R. Ramamoorthi]
{\parbox{\textwidth}{\centering         Nithin Raghavan,$^{1*}$
      Yan Xiao,$^{1*}$
      Kai-En Lin,$^{1}$
      Tiancheng Sun,$^{2}$
      Sai Bi,$^{3}$
      Zexiang Xu,$^{3}$
      Tzu-Mao Li,$^{1}$
      Ravi Ramamoorthi$^{1}$
        }
        \\
{\parbox{\textwidth}{\centering       $^1$University of California, San Diego\quad
    $^2$Google\quad
    $^3$Adobe Research\\
    \textit{$^*$ denotes equal contribution}
       }
}
}

\begin{document}


\maketitle
\begin{abstract}
   Precomputed Radiance Transfer (PRT) remains an attractive solution for real-time rendering of complex light transport effects such as glossy global illumination.  After precomputation, we can relight the scene with new environment maps while changing viewpoint in real-time.  However, practical PRT methods are usually limited to low-frequency spherical harmonic lighting.  All-frequency techniques using wavelets are promising but have so far had little practical impact.  The curse of dimensionality and much higher data requirements have typically limited them to relighting with fixed view or only direct lighting with triple product integrals. In this paper, we demonstrate a hybrid neural-wavelet PRT solution to high-frequency indirect illumination, including glossy reflection, for relighting with changing view.  Specifically, we seek to represent the light transport function in the Haar wavelet basis.  
    For global illumination, we learn the wavelet transport using a small multi-layer perceptron (MLP) applied to a feature field as a function of spatial location and wavelet index, with reflected direction and material parameters being other MLP inputs. We optimize/learn the feature field (compactly represented by a tensor decomposition) and MLP parameters from multiple images of the scene under different lighting and viewing conditions.  We demonstrate real-time (512 x 512 at 24 FPS, 800 x 600 at 13 FPS) precomputed rendering of challenging scenes involving view-dependent reflections and even caustics.

\ccsdesc[300]{Computing methodologies~Reflectance modeling}

\printccsdesc   
\end{abstract}  
\section{Introduction}
Interactive rendering of scenes with complex global illumination
effects remains a long-standing challenge in computer graphics.
Precomputed Radiance Transfer (PRT)~\cite{Sloan1}, which enables interactive 
relighting by precomputing the light transport of a static scene, 
remains an attractive solution.  However, the
practical impact of PRT has largely been limited to low-frequency
spherical harmonic methods.  All-frequency methods using Haar wavelets
were proposed to address this shortcoming, but required substantially
larger data storage, and were therefore limited to fixed
viewpoint~\cite{Ravi2}, triple products for direct lighting
only~\cite{Ravi3} or lower-frequency BRDF in-out
factorizations~\cite{Sloan4,RuiWang}.  Obtaining true all-frequency
relighting with changing view-dependent glossy global illumination effects
requires precomputing, storing and rendering with a high-resolution 6D
light transport tensor for spatial, light and view variation, which
has remained intractable because of the exponential growth in data
size with dimensionality.

With the advent of deep learning and implicit neural representations,
we have a new mathematical toolbox of function approximators that can
be used to revisit this challenge.  Indeed, work on neural radiance
fields~\cite{nerf,Tancik} showed that high-dimensional spatio-angular
radiance functions can be learned by a simple multi-layer perceptron
(MLP), and these ideas have been applied to directly solve the
rendering equation with neural function representations~\cite{neuralradiosity}.
However, simply approximating the light transport matrix in a neural
basis is insufficient for PRT, since one needs to compute light
transport integrals in real-time as is done in spherical harmonics or wavelets.

In this paper, we leverage the seminal early PRT work, modern
MLP-based function approximators and recent rendering advances to
tackle these problems.  We focus on indirect lighting, including
glossy view-dependent global illumination. 
Several approaches to real-time direct
lighting exist, including the original triple product
formulation~\cite{Ravi3}, and ReSTIR~\cite{restir}.  
We leverage real-time raytracing in OptiX on modern RTX
graphics cards~\cite{nvidialatest,Optix} with importance sampling of
the environment map and Monte Carlo
denoising~\cite{ZwickerSTAR} in OptiX~\cite{optix5}.  However, such a direct path-tracing
approach is still not real-time for complex light transport paths
involving multi-bounce glossy indirect reflections or caustic
patterns.  Our major technical contributions and design decisions include:

\begin{figure*}
    \centering
    \includegraphics[width=1.0\textwidth]{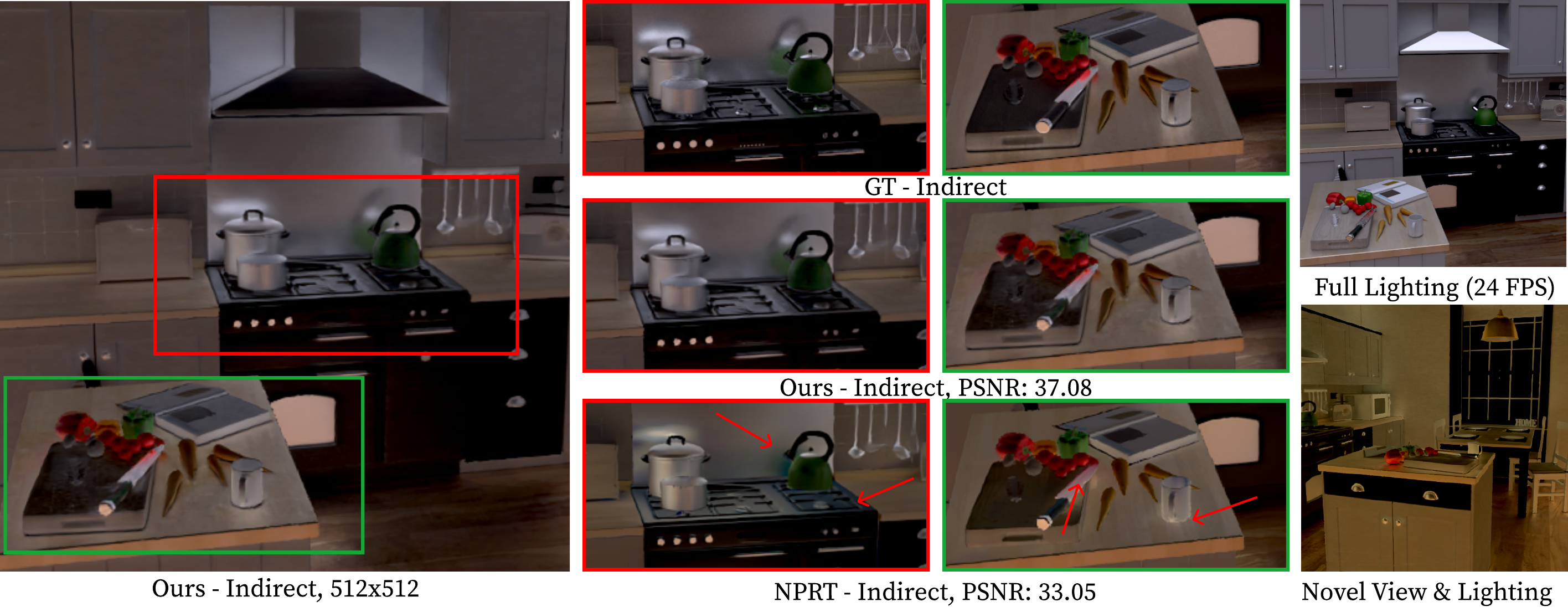}
    \caption{We develop a precomputed radiance transfer (PRT) method, based on a hybrid neural-wavelet representation. Our method enables high-frequency relighting with changing view and glossy indirect illumination.  {\em Left:}  Indirect illumination (which our method focuses on) rendered at 24 FPS on an RTX 4090 with our system using 64 Haar wavelets for the environment map and our learned MLP light transport. {\em Middle: } Comparison to Neural PRT~\cite{neuralprt} and ground truth.  Neural PRT does not handle high-frequency view-dependent effects as well as our method (notice the missing glossy reflections pointed out by the arrows), and has a slight tonal shift on the stove and the pots. {\em Right: } Showing the rendering combined with direct lighting, and different lighting environments and views of the same scene rendered in real-time.}
    \label{fig:teaser}
\end{figure*}

\vspace*{-.05in}
{\em Haar Wavelet Representation: } As in the
original wavelet-based PRT algorithms, we seek to project the lighting
and light transport into Haar wavelets, while keeping a small number
(typically 64) of the most important lighting coefficients.  This
enables a real-time wavelet dot product and projection of the
environment map as in previous work, and differs from recent neural
PRT approaches~\cite{neuralprt,lightweightneural}, which require
separate neural layers to compute dot products within neural
functions.
While Rainer et al.'s method~\shortcite{neuralprt} is suitable for largely diffuse scenes, the quality of indirect view-dependent effects is often less accurate.
Their neural approximation of the linear dot product can also lead to a tonal shift of the image.
By working directly with wavelets, our approach better preserves high-frequency effects such as glossy reflections, and it has the theoretical benefits in remaining grounded in linear function spaces (see Fig.~\ref{fig:teaser} and quantitative comparisons in Table~\ref{tab:quantitative}). 

\vspace*{-.1in}
{\em Light Transport Representation: }
The key challenge is the represention of the light transport
coefficient for a given view, spatial location and wavelet index.  For
direct lighting, the 6D light transport can be factorized into
a product of 4D view-independent visibility and 4D BRDF functions,
with wavelet coefficients of the product computed using triple
products~\cite{Ravi3}.  However, it is not possible to extend this
formulation to global illumination.  We make two important
modifications, enabled by modern MLP-based learning algorithms.
First, instead of visibility, we learn a feature vector parameterized
by spatial location and wavelet index.  To enable compact storage and
fast training and evaluation, we decompose the feature field with
tensor factorization~\cite{Chen2022ECCV}, where the 3D spatial component is
represented using a multiresolution hash grid~\cite{instantNGP}.  To
our knowledge, this is the first method to combine the use of tensor
factorization and multiresolution hash grids.  Finally, we use a 
small MLP that takes as input the feature vector, reflection 
direction, normal, and BRDF parameters, and outputs the transport 
coefficients.  The MLP and feature field tensors are all trained on 
images of the scene under different views and environment maps.

\vspace*{-.05in}
{\em Real-Time Rendering: } 
We demonstrate real-time rendering with precomputed light transport,
including glossy effects with changing lighting and view.  The
size of our final representation is compact (113 MB for the scene in Fig.~\ref{fig:teaser}), significantly smaller than an explicit 6D representation, and even smaller than 
early wavelet-based relighting methods without view-dependence~\cite{Ravi2}.
We believe
our work addresses a long unresolved challenge in PRT methods, to
enable high-frequency lighting and viewpoint variation with global
illumination (see Fig.~\ref{fig:teaser}, rendered at 24fps), while giving a new capability to real-time rendering.


\section{Related Work}
PRT research has always relied on new mathematical representations
beyond spherical harmonics and wavelets, such as zonal
harmonics~\cite{Sloan5}, clustered principal components
(CPCA) \cite{Sloan2}, spherical Gaussians with tensor
approximation~\cite{Tsai}, and von-Mises Fisher approximations of the transfer function~\cite{ruigather}.  Our work can be seen as a natural
progression along this line involving MLP-based neural function
approximators.  We limit ourselves to the standard PRT
setting of static scenes with distant environment map illumination,
and do not consider near-field area sources~\cite{Mei}, or dynamic
objects~\cite{Shum}. We are distinct from direct-to-indirect transfer
methods~\cite{Pellacini2, datadrivenprt}, which cannot easily handle complex
view-dependent global illumination. We do take inspiration from them
in handling direct lighting separately.  We refer readers to the
comprehensive survey by Ramamoorthi ~\cite{prtsurvey}, which points out
the unsolved nature of all-frequency relighting with changing
viewpoint and glossy objects.  They also note that triple
products~\cite{Ravi3} are limited by the inability to support spatial
compression (CPCA), while BRDF factorization
methods~\cite{Sloan4,RuiWang} can require more than a hundred terms
for high-frequency materials~\cite{Dhruvegsr}.

Ren et al.~\shortcite{PeiranRen2,PeiranRen1} first introduced the use
of neural networks to regress global illumination and image-based
relighting.  In contrast, we focus on the classic PRT problem with
environment maps, and introduce a novel light
transport representation.  
Most recently, Rainer et al.~\shortcite{neuralprt} introduced a neural PRT solution with diffuse-specular separation.  They do not directly use
wavelets, unlike our method, but use a convolutional neural network 
to extract lighting features.  In contrast, our method is a novel
hybrid using neural networks to predict Haar wavelet transport
coefficients, and we demonstrate better glossy
effects in our results, and better quantitative metrics (see Table~\ref{tab:quantitative} in results). 

Xu et al.~\shortcite{lightweightneural} introduce lightweight neural
basis functions, and neural networks for double and triple product
integrals. As with most neural bases, there is no guarantee of
orthonormality and a separate network is needed for the dot products.
In contrast, we leverage standard orthonormality and approximation with the most significant coefficients by
performing dot products in wavelets, while using neural networks only
to represent wavelet transport coefficients.  Moreover, Xu et al. only
demonstrate fixed view, and the inherent limitations of triple product
integrals require a restriction to direct lighting.  Our work also
relates to research on neural materials and layering~\cite{nlbrdf} and
recent efforts in acquisition of light transport from real
scenes~\cite{nrtfields} but we have very different goals.

We acknowledge the significant recent progress in real-time path
tracing and denoising~\cite{svgf,learn2} without the need
for any precomputation.  A comprehensive discussion of these methods
is out of scope, and they are largely orthogonal to our PRT-based
approach.  We do note that they are usually still limited in capturing
complex multi-bounce light transport like glossy reflections at the
low sample counts required for real-time applications.  We do leverage
this research by denoising the direct lighting.  Although our PRT indirect renderings are of high quality, and not affected by Monte Carlo noise in the traditional sense, we do observe a small benefit from denoising, see Table~\ref{tab:quantitative} and Fig.~\ref{fig:denoising}.  


\section{Overview}

We now provide a brief overview of our method.  The light transport
equation is given by,
\begin{equation}
B(\bm x, \bm \omega_o) = \int_\Omega T(\bm x, \bm \omega, \bm
\omega_o) L(\bm \omega)\,d\bm\omega,
\label{eq:1}
\end{equation}
where $B$ is the outgoing radiance we seek, as a function of surface
position $\bm x$, and outgoing direction $\bm \omega_o$.
It is given by an integral of the environment map lighting $L$, a
function of incident direction $\bm \omega$, multiplied by the light
transport function $T$, which is a function of spatial location $\bm
x$ and incident and outgoing angles $(\bm \omega, \bm \omega_o)$.  For
our purposes $T$ will represent only the global illumination
component, with direct lighting computed separately.  

In PRT, we precompute the light transport $T$ and
dynamically change the lighting $L$ and view $\bm \omega_o$.  We
follow previous PRT methods by projecting the lighting (at run-time)
and transport (precomputed) into a suitable basis---Haar wavelets on
cubemap faces as in previous work~\cite{Ravi2} 
\begin{eqnarray}
L(\bm \omega) & = & \sum_j L_j \Psi_j(\bm \omega) \\
T(\bm x, \bm \omega, \bm \omega_o) & = & \sum_k T_k(\bm x, \bm
\omega_o) \Psi_k(\bm \omega), \nonumber
\label{eq:lighting}
\end{eqnarray}
where $\Psi_k$ are the basis functions indexed by $k$, and $L_k$ and
$T_k$ are the lighting and transport coefficients.  The
basis expansion in $T$ is only over the incident direction $\bm
\omega$ which is being integrated.  We achieve real-time
rendering simply by taking the dot-product, 
\begin{eqnarray}
B(\bm x, \bm \omega_o) & = & \sum_j \sum_k L_j T_k(\bm x, \bm
\omega_o) \int_\Omega \Psi_j(\bm \omega) \Psi_k(\bm \omega)
\,d\bm\omega \nonumber \\
& = & \sum_{k \in K} L_k T_k (\bm x, \bm \omega_o) \nonumber \\
& = & \bm L \cdot \bm T(\bm x, \bm \omega_o),
\label{eq:dotproduct}
\end{eqnarray}
where $\bm L$ and $\bm T$ represent vectors of all coefficients $k$,
and the integral simplifies by the orthonormality of basis functions.
This simplicity and the resulting practicality for real-time
rendering is not possible when using a (non-orthonormal) neural basis
as in earlier work~\cite{neuralprt,lightweightneural}. These works
therefore require a separate more complex network to perform
approximate integration/dot products.  Efficiency in the summation or
dot-product is obtained by considering only a set $K$ of the largest
wavelet coefficients in the lighting (we
typically use 64); this is indicated in the second line above.  The
entire transport $T$ must still be precomputed, but only the
coefficients in $K$ will be used (this set can change at each frame with the lighting).

It remains to compute and represent $T$ and $T_k$.  As motivation, we first 
review the triple product approach used for direct lighting~\cite{Ravi3}.
In that case, the transport is simply the point-wise product of
(view-independent) visibility $V$ and cosine-weighted BRDF $\rho$, with wavelet coefficients 
computed using triple products, 
\begin{eqnarray}
T^d(\bm x, \bm \omega, \bm \omega_o) & = & V(\bm x, \bm \omega) \rho(\bm
\omega, \bm \omega_o) \nonumber \\
T^d_k(\bm x, \bm \omega_o) & = & \sum_i \sum_j C_{ijk} V_i(\bm x) \rho_j(\bm \omega_o) \nonumber \\
B^d(\bm x, \bm \omega_o) = \sum_k L_k T^d_k(\bm x, \bm \omega_o) & = & \sum_i \sum_j \sum_k C_{ijk} V_i(\bm x) \rho_j(\bm \omega_o) L_k \nonumber,
\end{eqnarray}
where $C_{ijk}$ are the tripling coefficients and we use the
superscript $d$ to specify this is for direct lighting only (these
equations are not used in our system; they are for illustration 
and motivation only).  Note that the original triple product 
method directly used the integration with the lighting (last line 
above) without explicitly forming the transport coefficient 
above, but this formulation is equivalent.

For global illumination, no such simple form exists and we will
represent $T_k(\bm x,\bm \omega_o)$ instead by a neural network.
However, we are inspired by the formulation above and modify it in two
key ways.  
First, as there is no closed-form expression for the convolution of visibility terms for an arbitrary number of ray bounces, we replace the visibility in the above formulation with a view-independent general feature
vector, which is a function of output wavelet coefficient $k$ and spatial
position $\bm x$.  This promotes a
compact factorization of light transport that allows the network to learn these terms.  Second, we replace the
simple multiplication of visibility and BRDF (and related triple
product wavelet formulation) by a small multi-layer perceptron (MLP)
that takes as input the feature vector, surface normal, reflected
direction and BRDF parameters (diffuse and specular coefficients,
roughness) and outputs the transport coefficient $T_k$. We provide the mathematical details in the next section.  

\section{Mathematical Framework}\label{sec:framework}

We now present a hybrid wavelet-neural framework, where transport
is computed in the wavelet basis as in the classical works, but
transport coefficients are determined by a neural network.  Regression
directly in the wavelet basis has several advantages. It is
well-established that the discrete transport operators are sparse in
the wavelet domain, as most of the frequencies are concentrated in
relatively few entries. This makes the problem of memorizing the light
transport for a particular scene tractable. Second, we can compute the
rendering equation directly, avoiding the need for
low-frequency approximations or using neural networks as renderers.
This allows for both view and lighting variations, enabling full
generalization for complex light transport effects.

{\em Representing Light Transport:}
Specifically, we represent the transport coefficients as, 
\begin{equation}
T_k(\bm x, \bm \omega_o) = f\left(\bm h_k(\bm x), \bm \omega_r(\bm x), \bm
  n(\bm x), \bm \rho(\bm x) ; \bm \Theta \right),
\label{eq:MLP}
\end{equation}
where $\bm h_k$ is a feature vector as a function of spatial
coordinate $\bm x$ and wavelet index $k$.  The feature field $\bm h$ in
essence captures how a wavelet light $k$ scatters with global
illumination when encountering scene point $\bm x$.  $f$ is a small
multilayer perceptron (MLP) network that decodes the feature vector
$\bm h_k$ into the appropriate light transport wavelet coefficient
$T_k$.  Additional inputs to the MLP are the reflected direction $\bm
\omega_r$, the reflection of the outgoing direction $\bm \omega_o$
about the surface normal $\bm n(\bm x)$, all in global coordinates. It is well known that using $\bm
\omega_r$ instead of $\bm \omega_o$ enables more coherent functions
that are easier to optimize/regress for~\cite{freqenv,refnerf}.  We
also pass in the BRDF parameters which we denote as a vector $\bm
\rho$, which could be spatially-varying.  We adopt a standard GGX/Trowbridge-Reitz
reflection model~\cite{Walter07,Trowbridge:1975:AIR}, with parameters $\bm \rho$ including
the diffuse and specular colors $\bm k_d$ and $\bm k_s$ and roughness $\sigma$.
$\bm \Theta$ denotes the parameters of the MLP.

{\em Feature Field Representation: } 
We have so far considered the feature vector $\bm h_k(\bm x)$ for a
given wavelet index $k$ and spatial point $\bm x$.  For compact
representation, it is convenient to explicitly write the feature field
$\bm h$ as a tensor $H$ with explicit parameters/indices, (we use notation $[]$ for accessing feature grids and $()$ for functions)
\begin{equation}
\bm h \equiv H[\bm x, \bm k, l],
\end{equation}
where spatial location is designated by $\bm x$ as before, a 3D
vector.  It is convenient for later representation to view the wavelet
index as a 2D vector $\bm k$, corresponding to position on a cubemap
after non-standard Haar wavelet decomposition.  Finally, $l$ is the 1D
index of the feature vector (typically we use a vector of length 64).  Note that explicitly representing
$H$ can be prohibitive, given it is effectively a 6D tensor.
Therefore, we develop a compact tensor factorization, inspired by
previous tensor approximations in the PRT and NeRF
literature~\cite{Tsai,Chen2022ECCV}.  This approach also has
similarities to PCA matrix decompositions, although we use a
multiresolution hash grid~\cite{instantNGP} for further compression
rather than clustering as in previous PRT works~\cite{Ko2,Sloan2,Tsai}.
Specifically, we use a CP tensor decomposition along the three modes
(spatial $\bm x$, wavelet $\bm k$, feature $l$) with $M$ terms to
write,
\begin{equation}
H[\bm x, \bm k, l] \approx \sum_{m=1}^{M} S_m[\bm x] W_m[\bm k] U_m[l].
\end{equation}
In the equation above, $S_m[\bm x]$ is itself a 3D spatial feature
grid depending on spatial coordinate $\bm x$, with
trilinear interpolation to obtain the value at any $\bm x$.  We
represent $S_m$ as a three-dimensional multiresolution hash
encoding~\cite{instantNGP} for ease of training and evaluation.
This differs from most previous works that store information on 
scene vertices.  
In our experiments, we found that such a volumetric representation results in fewer view-dependent artifacts than a scene vertex representation (see Table \ref{tab:ablation_hashgrid}) or a learned neural texture (single-resolution hash grid), and is easier to implement and compress, since parameterization of geometry remains a difficult problem.
Note that the rendering costs of volumetric methods are independent of the level of detail of the scene; this has been exploited in previous works involving neural scene-to-volume computation \cite{bakoprefiltering}. 
$W_m[\bm k]$ is a two-dimensional grid that stores a feature vector for each wavelet.  Since the environment map is represented with a cubemap, wavelets and $W_m$ can also be represented as a cubemap.
Finally, $U_m[l]$ represents the ``feature''
dimension, which is a 1D vector for each $m$, where
$\bm U$ itself is simply a learnable matrix.  

Given the tensor decomposition of the feature field, we can evaluate
the feature vector in Equation~\ref{eq:MLP} at runtime as follows, 
\begin{equation}
\bm h_k(\bm x) = \sum_{m=1}^{M} S_m[\bm x] W_m[\bm k] \bm U_m, 
\label{eq:featurevector}
\end{equation}
where $\bm U_m$ denotes the vector corresponding to all $l$ in $U_m[l]$.  

\begin{algorithm}[!!t]
\begin{algorithmic}[1]
\Procedure{Render}{$L$,$\{S_m\}$,$\{W_m\}$,$\{\bm U_m$\},$M$,$\bm \Theta$} 
\State $L \rightarrow L_k$ \label{line2} \Comment{Wavelet Transform Lighting, Equation~\ref{eq:lighting}}
\State $K \rightarrow$ set of largest wavelet coefficients $L_k$ \label{line3}
\ForAll{pixels}
\State Determine $\bm x$, $\bm \omega_r$, $\bm n$, $\bm \rho$ \label{line5} \Comment{Inputs to Equation~\ref{eq:MLP}} 
\For{$k \in K$} 
\State $\bm h_k = \sum_{m=1}^{M} S_m[\bm x] W_m[\bm k] \bm U_m$ \label{line8} \Comment{Equation~\ref{eq:featurevector}}
\State $T_k = f\left(\bm h_k, \bm \omega_r, \bm n, \bm \rho ; \bm \Theta\right)$ \label{line9} \Comment{MLP in Equation~\ref{eq:MLP}} 
\EndFor
\State $B = \sum_{k \in K} L_k T_k$ \label{line10} \Comment{Equation~\ref{eq:dotproduct}}
\State $B = B + B^d$ \label{line12} \Comment{Add denoised direct lighting}
\EndFor
\EndProcedure
\end{algorithmic}
\caption{PRT Rendering}
\label{fig:prtrender}
\end{algorithm}

{\em High-Level Rendering Algorithm: } 
Algorithm~\ref{fig:prtrender} shows the pseudocode of our global illumination rendering algorithm.  We first decompose the environment
map into wavelets (Equation~\ref{eq:lighting}, line~\ref{line2}) and
pick the set of largest wavelet coefficients $K$ (line~\ref{line3}).  The feature field
$\bm h \equiv \{S_m,W_m,\bm U_m\}$ is stored and learned compactly using a tensor decomposition
and multiresolution hash grid, as discussed above.  For a given pixel,
we use rasterization or raytracing to find the primary hit at a pixel,
with spatial location $\bm x$, normal $\bm n$, outgoing/reflected
directions $\bm \omega_o$ and $\bm \omega_r$ and BRDF parameters $\bm
\rho$ (line~\ref{line5}).  

Now, for each wavelet index $k \in K$, we determine the
feature vector $\bm h_k(\bm x)$ (see Equation~\ref{eq:featurevector}, line~\ref{line8}).
We now evaluate the MLP $f(\cdot)$ in Equation~\ref{eq:MLP} (line~\ref{line9}) to obtain
the transport coefficient $T_k$.  Once the vector of all transport
coefficients $T_k$ with $k \in K$ is obtained, we determine the final
color by performing the dot product with lighting in
Equation~\ref{eq:dotproduct} (line~\ref{line10}). We also add in the denoised direct lighting, computed separately (line~\ref{line12}).  

\section{Implementation and Algorithm}
We now proceed to discuss the implementation and algorithm, based 
on the mathematical framework in the previous section.

\begin{figure*}[!htb]
\centering
    \includegraphics[width=1.0\textwidth]{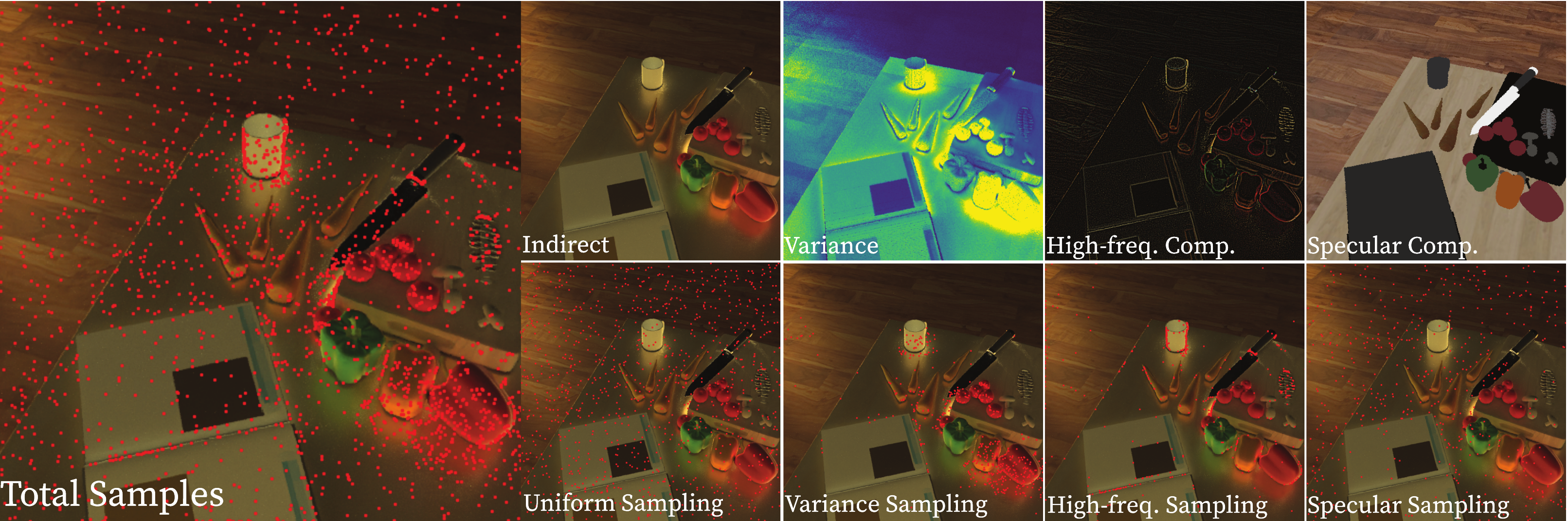}
    \caption{Sampling strategy used for precomputation/learning of indirect light transport. The left image shows the total sample point distribution. The points are made up of uniform samples over the image, and concentrations of samples in regions of high view-variance, high-frequency regions, and specular materials.}
\label{fig:sampling}
\vspace*{-.2in}
\end{figure*}

\vspace*{-.05in}
{\em Precomputation: Rendering.} 
As with all PRT algorithms, there is a precomputation step.  In our
case, this involves not only offline rendering of the scene of
interest, but also learning the relevant light transport parameters.
We use a GPU-accelerated path tracer in OptiX with
denoising to produce 512x512, 1024 samples per pixel ground truth images for training.  
Each  image takes 1-3 seconds to render and it is not interactive, underscoring the
need for our real-time PRT algorithm.  The image resolution for real-time rendering can be changed arbitrarily at run-time, and we use higher-resolution $800\times 600$ renders in some of our results. 

For a given scene, we render
approximately 4000 images under different environment maps and viewing
conditions.
We use 1000 indoor cubemaps and rotate each by 120 and 240 degrees to obtain the 3000 training lighting conditions. We only select indoor ones instead of outdoor ones since nonlinear wavelet selection on those tends to result in a larger quantity of meaningful wavelet coefficients \cite{Ravi2}. We generate 2000 camera locations using trajectories placed in the scene, and for each camera, we randomly select 2 environment maps from our training pool. 
We use one-sample-per-pixel raycasting to obtain the geometry parameters, reflection direction and BRDF parameters for these training views.
This precomputation step takes about 1-3 hours.  Note that the number of images is almost an order of magnitude less than the number needed in early wavelet methods, even for fixed view~\cite{Ravi2}. 


We found that for highly specular areas, the algorithm requires multiple samples of view-dependent effects under different lighting conditions. For simple scenes (\textsc{Four Animals}) where the camera can see almost every object at a given time, we place the cameras on predetermined spherical trajectories. For scenes that have many occluded areas (\textsc{Kitchen} and \textsc{Armadillo}) we add an additional helical trajectory. 

\vspace*{-.05in}
{\em Precomputation: Learning Light Transport.} 
The trainable parameters in our formulation are the feature grids
$\{S_m\}$, $\{W_m\}$ and $\{\bm U_m\}$ as well as the parameters for
the MLP $f$, which we denote as $\bm \Theta$. In particular, $\{S_m\}$ is represented as a multiresolution hash grid, which concatenates features obtained by trilinear interpolation across resolutions. 
Though past PRT methods have generally stored the feature vectors representing exitant radiance densely along the vertices of a mesh, we found that using such a volumetric representation significantly improves performance (see Table~\ref{tab:ablation_encodings}). 
$\{W_m\}$ is represented as a neural texture at the same resolution as the cubemap, $6\cdot 64\cdot 64$. 
We set the number of terms $M$ for both feature grids to be $64$, which we found gives the best tradeoff between accuracy and speed, and we also set the feature dimension of the hash-grid to be $64$ (so $\bm U$ becomes a square matrix) as we found reducing this value does not meaningfully reduce the computation time. For ease of implementation, the learnable matrix $\bm U$ is represented as a single-layer fully-fused MLP with no bias. $f$ is implemented as a two-layer fully-fused MLP with width 128. The total size of our precomputed
data is about 113 MB, the bulk of which stores the 3D multiresolution hashgrid representing
$\{S_m\}$. This is substantially less than previous methods~\cite{Ravi2,Ravi3}
even though we are considering full 6D indirect light transport.
Our goal is to optimize
\begin{equation}
\{S_m\}, \{W_m\}, \{\bm U_m\}, \bm \Theta = \mathrm{arg\,\,min}\
\mathcal{L}\left(I\left(S_m, W_m, \bm U_m, M, \bm \Theta\right),I_0\right),
\end{equation}
where $I$ is the image rendered using the procedure in
Algorithm~\ref{fig:prtrender} and $I_0$ is the ground truth rendering discussed above.

At each training step, we randomly select an environment map from the training data, perform the non-standard wavelet decomposition over cubemap faces as in \cite{Ravi2} and select 300 wavelets. The choice of 300 is motivated by past findings (\cite{Ravi2}, \cite{Ravi3}) noting that over 99\% $L^2$ accuracy can be obtained by choosing less than 1\% of the total wavelets. We importance sample half of these wavelets via unweighted selection from the environment map, and as the largest entries of the ground-truth wavelet transport matrix are uncorrelated with such a purely top-$k$ selection, we uniformly sample wavelet indices for the other half to form $K$ and $L_k$.
We found that performing the wavelet transform without premultiplying the environment map entries by their solid angle factors (in effect, allowing the network to learn these) tends to produce better results.

We then sample 2048 pixels from the subset of our training data corresponding to this environment map and pass them through our algorithm to obtain the wavelet coefficients $T_k$ corresponding to the indices $k$, which we multiply with $L_k$ to obtain our final rendering.
The network tends to converge much more slowly on the highly view-dependent areas of the scene, so we adopt a specialized importance sampling strategy on these pixels (see Fig.~\ref{fig:sampling}).
In addition to the geometry buffer, we compute the empirical variances of all the hit points of the scene (stored in half-precision) and the high-frequency regions (obtained by subtracting a low-pass filtered version of the ground-truth indirect illumination from the original image). 
To deal with moderate-frequency regions we also importance sample based on the product of the specular coefficient times the roughness complement $k_s \cdot (1 - \sigma)$, and deal with all other regions via standard uniform sampling. 
We treat the output of these strategies as a probability distribution and sample 512 pixels from each accordingly. 

We opt for such an image-based strategy 
as it is faster than supervision using the full ground-truth light-transport $T$. The latter, which would entail generating a 6D tensor at resolutions of $6 \times 64 \times 64$ for lighting and view (as used for cubemaps in \cite{Ravi2, Ravi3}), would require over $(6 \times 64 \times 64 )^2 \approx 6 \times 10^8$ images. Additionally, experiments showed that even if this tensor were subsampled at multiple views, the resulting convergence of the network was inadequate to getting good results on novel-view specularities.
In the future, a more adaptive active exploration approach may be helpful to increase the training time spent on hard-to-learn examples and prevent overfitting on the diffuse parts of the scene ~\cite{activexploration}. 

We now compute the loss. Past works have demonstrated that error from applying an $L2$ loss directly on output HDR images tends to be disproportionately affected by bright regions, so we apply a tonemap to our prediction and the ground-truth rendering before we take the loss. 
We use the $\mu$-law tonemapping \cite{Kalantari3} $\mathrm{TM}(x) = \mathrm{sgn}(x) \frac{\log\left(\mu\cdot |x| + \epsilon\right)}{\log\left(\mu+\epsilon\right)}$, with $\mu=10$ and $\epsilon=1$, and define the loss as $\mathcal{L}(I,I_0) = \Vert \mathrm{TM}(I) - \mathrm{TM}(I_0) \Vert_2^2$. The extension to the negative numbers is required as our network operates directly in the wavelet domain, so our initial network predictions may result in negative colors after multiplication. 

We discuss the resolutions and encoding of the feature grids and MLPs. 
For the three-dimensional multiresolution hash grid $\{S_m\}$, we use 32 levels, 2 features per level, base resolution 16, a hash table size of $2^{19}$ and a per-level scale of $1.3$. This takes up the bulk of the total size of our method at 107 MB. While choosing a smaller hash table would result in a smaller model size, we found that it corresponded to a significant decrease in performance; see Table~\ref{tab:hashmapsize} for ablations on different hash table sizes.
For the two-dimensional grid $\{W_m\}$, we store $6\cdot 64\cdot 64\cdot 64$ values as full-precision floats, which takes up 6.1 MB.
We represent the learnable matrix $\bm U$ as a single-layer neural network with no bias or nonlinearity, taking up 36 KB. 
The final MLP takes as input the normal, reflection direction, and BRDF parameters and encodes only the reflection direction with spherical harmonics of maximum degree $4$. The input roughness $\sigma$ is additionally mapped to $\frac{\log(25\sigma + 1)}{\log{(25+1)}}$ for better resolution in areas with low roughness. An evaluation of our encoding scheme can be found in Table~\ref{tab:ablation_encodings}. This final MLP has 128 neurons and 2 hidden layers with ReLU activations, resulting in a size of 124 KB. Further significant compression is possible using dithering and aggressive quantization of values (we
use floats).  The total training time is typically around 16 hours on an NVIDIA RTX 3090Ti. 



\vspace*{-.05in}
{\em Real-Time Rendering.} Our renderer is implemented in C++ with 
Optix, CUDA,
and Tiny-CUDA-NN~\cite{Muller2}.  As noted earlier, direct
lighting is computed separately by importance sampling the environment
map and denoising (other approaches could also
be used).  We compute indirect lighting per
Algorithm~\ref{fig:prtrender}. Tiny-CUDA-NN is used for our neural
rendering step to obtain the coefficients $T_k$ by evaluating the MLP $f$, 
and we have a final CUDA kernel 
to compute the dot product of the transport coefficients and the environment map wavelet coefficients. 

In practice, we have found a modest
benefit from denoising the indirect lighting as well to avoid wavelet
noise and we, therefore, apply denoising to the combined direct and
indirect (this takes only about 1.5 ms and adds minimal overhead).
However, our results remain of high quality even without denoising (see Fig.~\ref{fig:denoising}). 

\begin{figure*}[!!t]
\centering
   \includegraphics[width=0.95\textwidth]{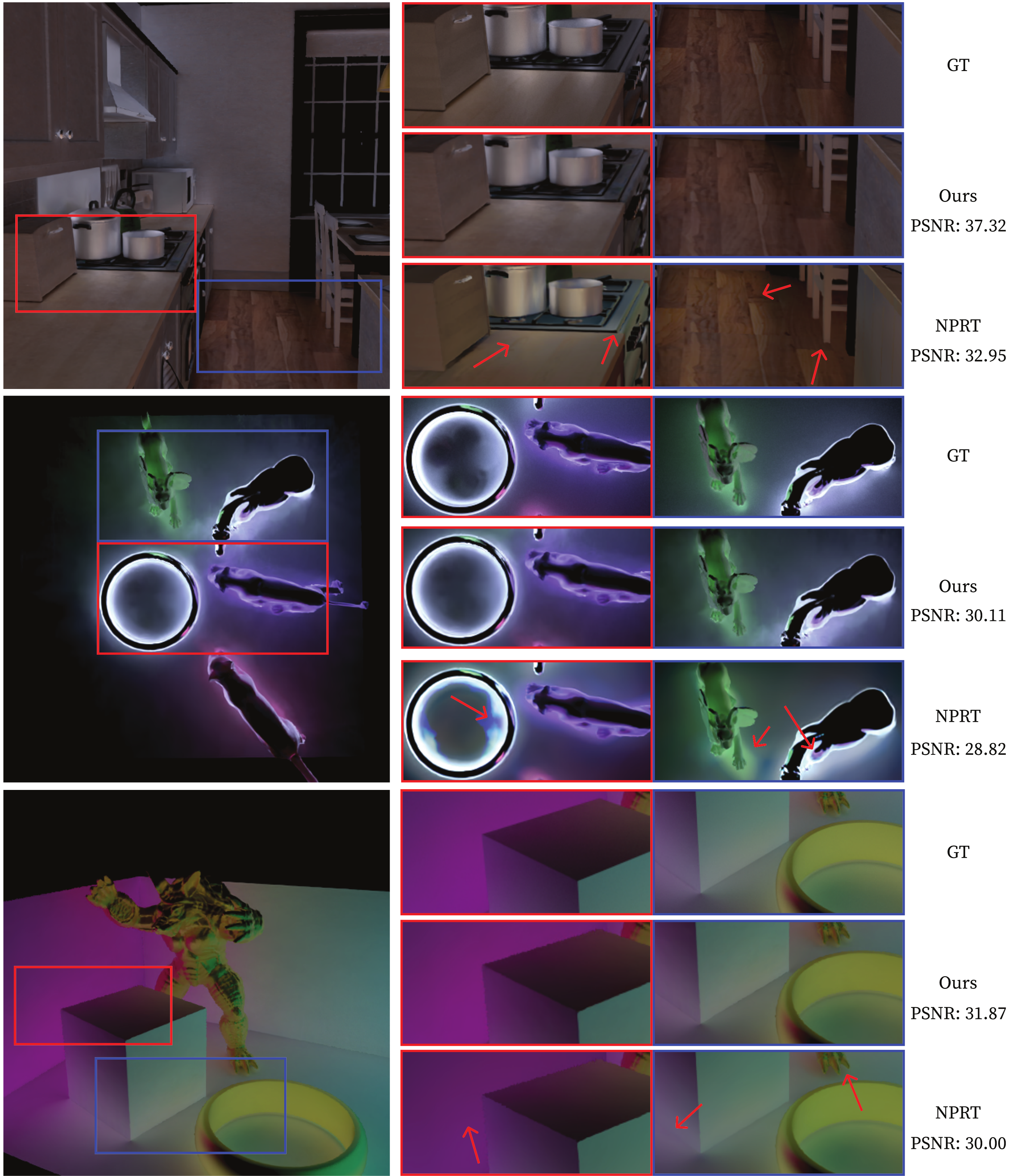}
\caption{
Visual comparisons of our method against Neural PRT~\cite{neuralprt},
Our method is able to reconstruct complex indirect illumination effects.
For example, see the glossy reflections on the table and the floor in the \textsc{Kitchen} scene (top row), the caustics in the \textsc{Four Animals} scene (mid row), and the color bleeding in the \textsc{Armadillo} scene (bottom row).
%
}
\label{fig:qualitative}
\end{figure*}

\section{Results}\label{sec:results}
We show results from our system, including 
screen-captures of the scenes we demonstrate (see video for real-time captures),
comparisons to previous work, and an evaluation of some of our parameters.
We compare to the best performing model described in Neural PRT \cite{neuralprt}, diffuse-specular separation. 

\subsection{Evaluation Methods}
We created our evaluation dataset of lighting and views for numerical comparisons using held-out environments and views. We selected 70 unseen indoor and outdoor environment maps. Our evaluation views include 
5 hand-picked locations that cover most objects in the scene and roughly 500 
locations generated using evaluation trajectories, which consist of helices, 
circular sweeps, or figure eights. We include videos generated using some of these evaluation trajectories in the supplementary material, and visual results show representative lighting environments and views. Note that PSNR numbers in Figs.~\ref{fig:teaser} and~\ref{fig:qualitative} refer to the scene with specific lighting/viewing configuration shown, and differ slightly from the averages over all configurations reported in Table~\ref{tab:quantitative}. For all the performance metrics (PSNR, SSIM, and LPIPS) reported in the tables, we show full direct+indirect / indirect only.

\subsection{Visual Results}
\begin{figure*}[t]
    \centering
    \includegraphics[width=0.8\textwidth]{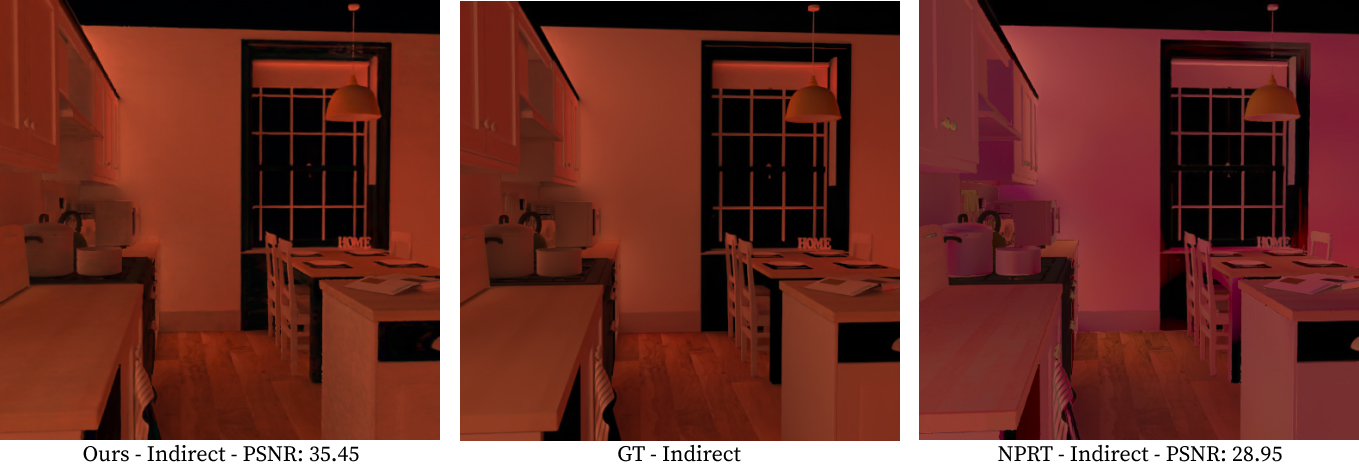}
    \vspace*{-.1in}
    \caption{Comparison of our method with NPRT on \textsc{Diffuse Kitchen} evaluated on a red environment map dissimilar to lighting conditions seen during training. In addition to an overall tone shift, the shading on certain objects (such as the pots on the stove) is inaccurate for NPRT, indicating our method is better able to generalize to unseen lighting conditions even for purely diffuse objects. }
    \label{fig:diffkitchen}
    \vspace*{-.1in}
\end{figure*}

\begin{figure*}[t]
    \centering
    \includegraphics[width=\textwidth]{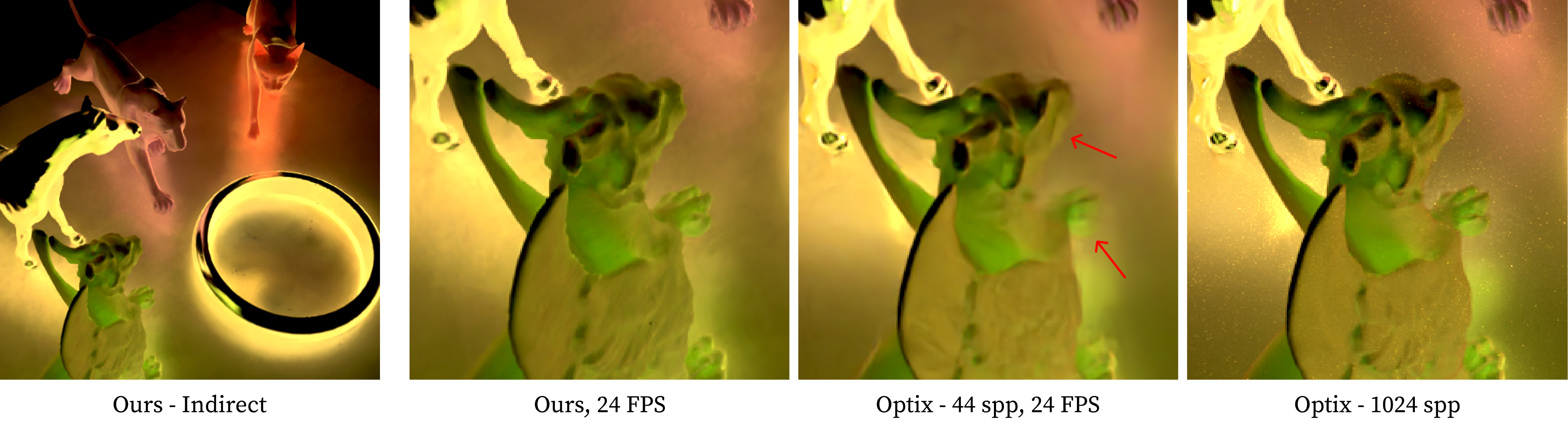}
    \vspace*{-.1in}
    \caption{Comparison of our PRT method against Optix path-traced renderings with the same rendering time (44 samples per pixel) under indirect lighting. Note the lack of high-frequency details around the head and the toes of the animal figurine in the OptiX render.}
    \label{fig:optix}
    \vspace*{-.1in}
\end{figure*}

In Figs.~\ref{fig:teaser} and~\ref{fig:qualitative}, we show example images from three scenes (all these results use denoising on both our method and the Neural PRT comparisons).  These are all rendered at 24fps for $512\times 512$ images and 13fps for $800\times 600$ images on an NVIDIA RTX 4090.  We see a range of
glossy indirect reflection effects, including view-dependent caustics
(see the \textsc{Four Animals} scene).  Capturing these high-frequency global illumination
effects has been very challenging in previous PRT algorithms, since
full high-resolution 6D lighting and view variation is required.
Our video clearly shows smooth motions for changing view and relighting. 
In Fig.~\ref{fig:teaser}, we see the \textsc{Kitchen} scene with glossy indirect reflections on the wall and the table.  Our method produces accurate indirect illumination and overall global illumination rendering.  In contrast, Neural PRT~\cite{neuralprt} works well for largely diffuse interreflections, but cannot perform well for high-frequency indirect highlights.
Additionally, the top row of Fig.~\ref{fig:qualitative} shows a different view of the \textsc{Kitchen} scene.
Neural PRT produces an overall browner tone compared to the reference, with a color shift on the stove.
It also misses glossy reflections on the table and the floor, making chair legs look flat and unnatural.  
Note that we retrained Neural PRT on each scene, using the same data as our method. 

We do not include comparisons with other methods, as traditional non-learning PRT approaches~\cite{Ravi2,Ravi3} are limited to fixed view or diffuse for double-products and direct lighting only for triple-products~\cite{lightweightneural}. Liu, et al.~\cite{Sloan4} and Wang, et al. \cite{RuiWang}'s approaches use the in-out BRDF factorization which is limited to lower-frequency reflections, while Hasan et al.'s method~\cite{Pellacini2} cannot render caustic paths and is similarly restricted to fixed view.


\begin{table*}[!!t]
  \centering
   {\caption{\em \textbf{Quantitative comparison of our method and Neural PRT (NPRT)}, both with and without denoising, for full direct+indirect / indirect only. These metrics are evaluated on views and lighting conditions both near and far from the training set. Our method is better on all metrics on all scenes.}
   \label{tab:quantitative}}
   \resizebox{\linewidth}{!}{
        {\begin{tabular}{l  l  c | c | c | c | c || c | c | c | c }  
        \toprule
        \multirow{2}{*}{Metric} & \multirow{2}{*}{Methods} & \multirow{2}{*}{Denoising} & \multicolumn{4}{c ||}{\textbf{Far Views}} & \multicolumn{4}{c}{\textbf{Near Views}} \\ 
        \, & \, & \, & Four Animals & Kitchen & Armadillo & Diffuse Kitchen & Four Animals & Kitchen & Armadillo & Diffuse Kitchen \\
        \midrule 
         \multirow{4}{*}{PSNR($\uparrow$)}            
         & NPRT & \xmark      
            & 28.08 / 24.35 & 33.80 / 31.43 & 34.53 / 30.92 
            & 43.20 / 39.65 
            & 28.23 / 24.44 & 35.05 / 32.08 & 34.30 / 30.89 & 43.63 / 40.10 \\               
         & NPRT & \cmark      
            & 28.53 / 25.09 &  34.56 / 32.45 & 34.74 / 31.53 & 43.45 / 40.55
            & 28.73 / 25.19 &  35.58 / 32.88 & 34.52 / 31.53 &  43.83 / 41.16 \\
         & Ours & \xmark      
            & 28.53 / 25.16 &  35.83 /  33.42 &  36.18 / 32.45 & 45.82 / 42.24
            & 28.77 / 25.38 &  36.71 / 34.93 & 36.59 / 32.68 &  46.18 / 42.68 \\
         & Ours & \cmark 
            &  \textbf{29.62} / \textbf{26.21} &  \textbf{36.59} / \textbf{34.31} & \textbf{36.48} / \textbf{33.17} & \textbf{46.27} / \textbf{43.69} 
            & \textbf{29.96} / \textbf{26.45} &  \textbf{37.64} / \textbf{35.59} & \textbf{36.94} / \textbf{33.44} &  \textbf{46.44} / \textbf{43.97}  \\
        \midrule
         \multirow{4}{*}{SSIM($\uparrow$)}                
         & NPRT & \xmark      
            & 0.9233 / 0.8408 & 0.9577 / 0.8957 & 0.9743 / 0.9445 & 0.9910 / 0.9540 
            & 0.9295 / 0.8550 & 0.9478 / 0.9128 &  0.9758 / 0.9483 & 0.9923 / 0.9607 \\  
         & NPRT & \cmark      
            & 0.9328 / 0.8527 & 0.9668 / 0.9165 & 0.9783 / 0.9511 & 0.9913 / 0.9620
            & 0.9394 / 0.8665 &  0.9683  / 0.9271 &  0.9799 / 0.9551 & 0.9927 / 0.9685 \\
         & Ours & \xmark      
            & 0.9064 / 0.8218 & 0.9647 / 0.9170 & 0.9750 / 0.9460 & 0.9945 / 0.9783
            &  0.9143 / 0.8374 & 0.9681 / 0.9211 &  0.9773 / 0.9512 & 0.9943 / 0.9765 \\
         & \textbf{Ours} & \cmark
            & \textbf{0.9390} / \textbf{0.8642} & \textbf{0.9767} / \textbf{0.9414} & \textbf{0.9822} / \textbf{0.9596} & \textbf{0.9947} / \textbf{0.9862}
            & \textbf{0.9441} / \textbf{0.8751} &  \textbf{0.9797} / \textbf{0.9434} &  \textbf{0.9840} / \textbf{0.9637} & \textbf{0.9950 }/ \textbf{0.9834} \\
        \midrule
         \multirow{4}{*}{LPIPS($\downarrow$)} 
        & NPRT & \xmark     
            & 0.0838 / 0.2049 & 0.0541 / 0.1792 & 0.0296 / 0.0851 & 0.0104 / 0.1009 
            & 0.0803 / 0.1948 &  0.0501 / 0.1665 &  0.0287 / 0.0844 & 0.0105 / 0.0921 \\  
        & NPRT & \cmark      
            & 0.0547 / 0.1788 & 0.0280 / 0.0970 & 0.0185 / 0.0559 & 0.0073 / 0.0593
            & 0.0496 / 0.1620 & 0.0275 / 0.1021 &  0.0187 / 0.0503 &  0.0071 / 0.0480 \\
        & Ours & \xmark     
            & 0.1315 / 0.1882 & 0.0498 / 0.1399 & 0.0293 / 0.0617 & 0.0059 / 0.0626
            & 0.1202 / 0.1774 &  0.0549 / 0.1408 &  0.0257 / 0.0609 &  0.0063 / 0.0621 \\
        & \textbf{Ours} & \cmark
            & \textbf{0.0489} / \textbf{0.1609} & \textbf{0.0161} / \textbf{0.0611} & \textbf{0.0121} / \textbf{0.0420} & \textbf{0.0029} / \textbf{0.0226}
            &\textbf{0.0453} / \textbf{0.1456} & \textbf{0.0211 }/ \textbf{0.0787} &  \textbf{0.0112} / \textbf{0.0364} & \textbf{0.0038} / \textbf{0.0277} \\
        \bottomrule
      \end{tabular}}
  }
\end{table*}

The middle row of Fig.~\ref{fig:qualitative} shows the \textsc{Four Animals} scene with challenging light transport with glossy reflection and a ring casting view-dependent caustics on a ground plane. These all-frequency view-dependent indirect reflections have historically been difficult for PRT methods, but our method produces fairly accurate results, where Neural PRT produces high frequency artifacts for the ring caustics and incorrect glossy reflections. Additionally, from the video of the \textsc{Four Animals} scene, we show that our method is also more temporally stable when it comes to rotation of high-frequency effects, while NPRT tends to be smoother with more incorrect shading.
Finally, the bottom row of Fig.~\ref{fig:qualitative} shows the \textsc{Armadillo} scene. Even in the largely diffuse regions, we still perform better than Neural PRT, since Neural PRT suffers from a color shift on the wall in the left inset and incorrect interreflections as indicated in the right inset.
Note the missing edge of the cube and the lack of indirect reflections on the ground from the claws.
The color shift of Neural PRT in the results above are likely due to the fact that Neural PRT does not use an orthonormal basis and has to approximate the linear dot product with a non-linear neural network.
To further investigate the behavior on diffuse reflections, we also include an almost entirely \textsc{Diffuse Kitchen} scene (all roughnesses set to $1$), shown in Fig.~\ref{fig:diffkitchen}.  We see that our performance is substantially better, because we do not suffer from color shifts or other network artifacts in computing wavelet dot-products.  

Finally, Fig.~\ref{fig:optix} makes a direct comparison of real-time low sample count (44 samples per pixel) path tracing with denoising in OptiX to our PRT rendering at equal time.  Note that this a highly favorable case for OptiX since the scene is geometrically simple, enabling a moderate brute-force path tracing sample count.  This sample count would be substantially lower in more complex production and game scenes.  Nevertheless, indirect reflections from the path tracer miss detail near the toes, head and wings of the animals, which are captured by our PRT rendering.  We have also observed greater temporal flickering in real-time path-traced renderings.

\begin{figure*} [!!t]
  \centering
    \vskip 0.12in
    \includegraphics[width=0.7\textwidth]{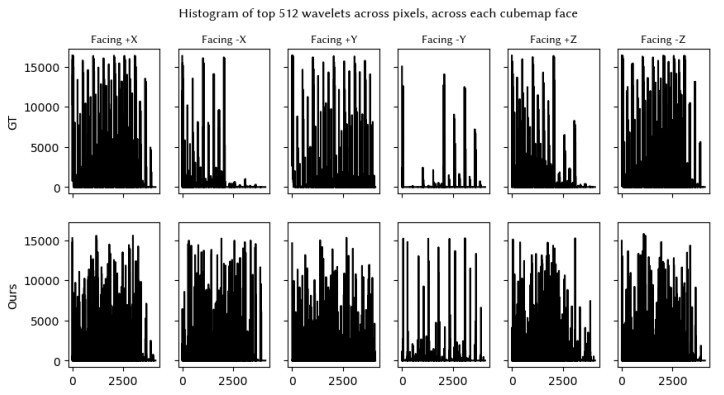}
    \caption{We fix a particular view of \textsc{Kitchen} from our far-views dataset and render out the ground truth wavelet transport matrix as in \cite{Ravi2}. We compare it to the equivalent transport matrix output via our method by visualizing the histogram of the top 512 wavelets over  all the pixels in the final image.}
    \label{fig:histogramandspread}
\vspace*{-.1in}
\end{figure*}

\begin{figure*} [!!t]
\begin{minipage}{.5\textwidth}
  \centering
    \includegraphics[width=\textwidth]{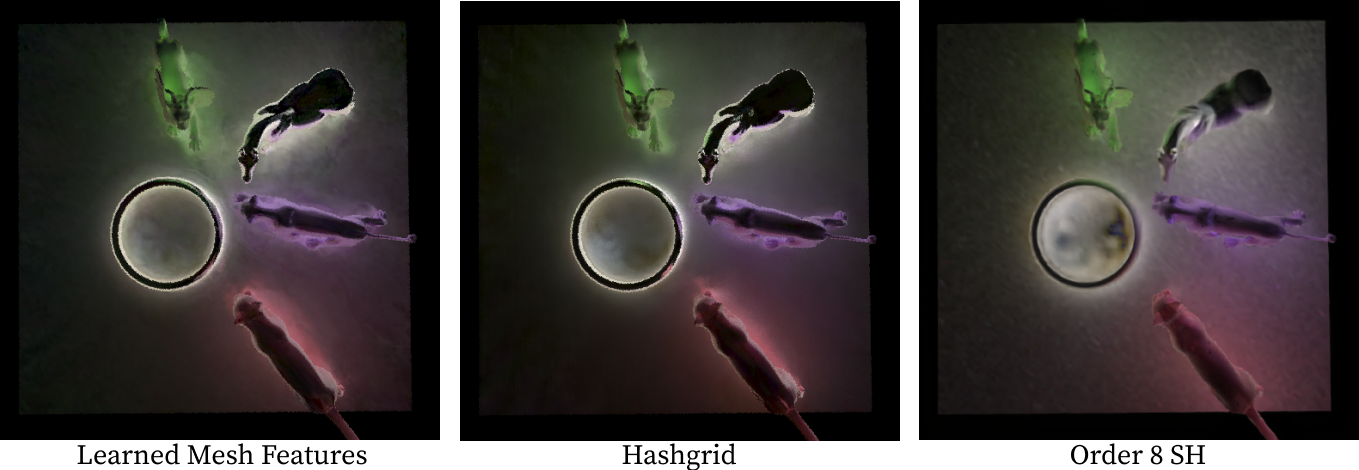}
    \vspace{-0.15in}
    \caption{Ablation study on storing the learned feature vectors on mesh vertices (learned and SH) vs a hashgrid. Spherical harmonics with a maximum degree of 8 stored on mesh vertices~\cite{Sloan1} are unable to properly represent caustics and the sharp reflections on the horse as expected, and also show some ringing and other artifacts. The learned mesh vertex features, while much better than SH, still cannot properly represent caustics, show a lot more noise in the reflections around the griffin and horse, and require a well-subdivided mesh. The use of the hashgrid solves these issues while being invariant to scaling of the scene.}
    \label{fig:representation}
\end{minipage}%
\hspace{0.11in}
\begin{minipage}{.5\textwidth}
  \vskip -0.45in
  \centering
  \includegraphics[width=1.0\textwidth]{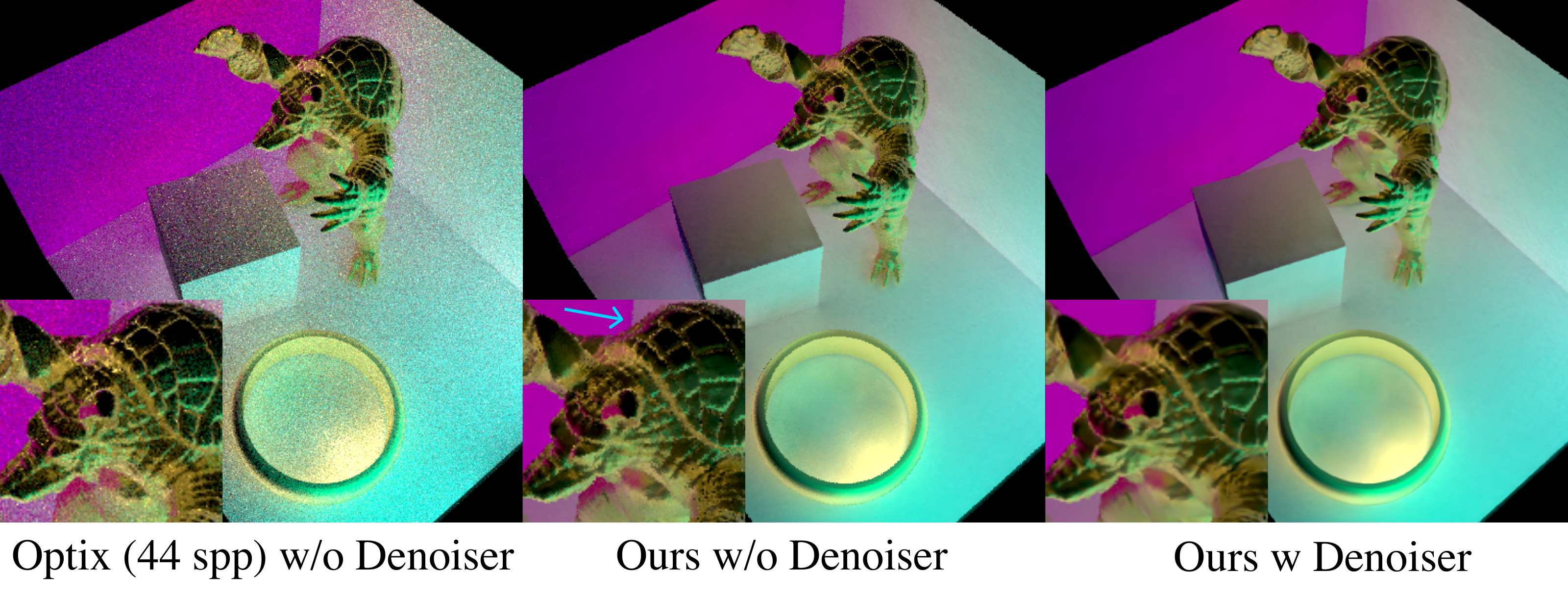}
  \vspace{-0.15in}
  \caption{The effects of the denoiser on our proposed method.  Standard OptiX path-tracing in equal time (left) is very noisy, and denoising does not address all issues (see Fig.~\ref{fig:optix}).  In contrast, our PRT rendering (middle) is already high quality without Monte Carlo noise, but does have a small amount of wavelet noise, and applying a denoiser helps improve the results (right) while adding almost no overhead.}
  \label{fig:denoising}
\end{minipage}
\vspace*{-.2in}
\end{figure*}

\subsection{Quantitative Results}\label{sec:quantitative}

Table~\ref{tab:quantitative} shows quantitative comparisons of Neural PRT and Our Method, both with and without denoising. 
We show results for both the full image and indirect lighting only on the metrics of PSNR, SSIM and LPIPS.
The main results on the left of the table are on novel lights and trajectories {\em far} from the training data, corresponding to the result figures, and showing the generalization ability.  The right of Table~\ref{tab:quantitative} shows additional statistics on held-out ({\em near}) views in training trajectories for both NPRT and our method.  

For all scenes and metrics, our method has significantly better accuracy than Neural PRT.
This is true both in scenes with strong glossy indirect reflections and complex caustics like \textsc{Kitchen} and \textsc{Four Animals}, and even when much of the global illumination is largely diffuse (\textsc{Armadillo} and \textsc{Diffuse Kitchen}).  
Both NPRT and our method have a small metric drop from training trajectories (near views) to novel ones (far views), and we perform better on all metrics in both cases.  This indicates our novel training strategy can generalize to unseen views and lighting conditions.

Our indirect PRT method does not have standard Monte Carlo noise, and denoising only provides a small (but important) boost.
Even without denoising, we are better than Neural PRT with denoising on the PSNR metric, and comparable on SSIM in most scenes.  Neural PRT does show a smaller bump in metrics after denoising than does our method; this is due to the fact that we predict coefficients directly in an orthonormal basis, which can result in noisier predictions in higher frequency regions. However, our method without denoising is also better than Neural PRT without denoising on almost all metrics.

Table~\ref{tab:performance} shows the runtime performance of our method on an RTX 4090.
Note that the performance is essentially the same for all scenes, independent of the scene's geometric complexity, because we use a volumetric hash-grid.
The rendering time for one frame scales approximately linearly with resolution ($512\times 512$ is about twice as fast as $800\times 600$) and the number of wavelets used (we used 64 wavelets for results in this paper, comparable to the number used in previous work~\cite{Ravi2,Ravi3}, but 32 wavelets may be suitable in lower-frequency scenes, with 128 wavelets needed in very challenging scenes for higher fidelity).
Rendering time remains interactive in all cases with real-time performance of 24fps or higher achieved for lower resolutions or number of wavelets.

\begin{table}[t]
  \centering
   {\caption{\em \textbf{Runtime Performance of our method with different resolutions, and \# of wavelets.}  We achieve interactivity in all cases.}
   \label{tab:performance}}
   \resizebox{\linewidth}{!}{
     {\begin{tabular}{l r r r} 
        \toprule
        Resolution     & \#Wavelets &  Framerate (FPS) &  Frame time (ms)    \\
        \midrule
                       &         32 &               42 & 24\\
        $512\times512$ &         64 &               24 & 42\\
                       &        128 &               12 & 87\\
        \midrule
                       &         32 &               25 & 40 \\
        $800\times600$ &         64 &               13 & 78 \\
                       &        128 &               6  &180\\
        \bottomrule
      \end{tabular}}
    }
\end{table}

\begin{table}[t]
  \centering
   {\caption{\em \textbf{Quantitative results of our model evaluated on \textsc{Four Animals} with different hashmap resolutions (direct+indirect/indirect only).} The results shown are after denoising. We find that a hashmap size of $2^{19}$ produces the best results in all metrics.}
   \label{tab:hashmapsize}}
   \resizebox{\linewidth}{!}{
     {\begin{tabular}{l r r r r} 
        \toprule
        HM size & Total size & PSNR ($\uparrow$) & SSIM ($\uparrow$) & LPIPS ($\downarrow$)   \\ \midrule
        $2^{17}$ & \textbf{34 MB} & 27.91 / 24.33 & 0.9219 / 0.8240 & 0.0687 / 0.1868 \\ 
        $2^{18}$ & 62 MB & 29.61 / 26.10 & 0.9370 / 0.8592 & 0.0508 / 0.1631 \\ 
        $2^{19}$ & 113 MB & \textbf{29.62} / \textbf{26.21} & \textbf{0.9390} / \textbf{0.8642} & \textbf{0.0489} / \textbf{0.1609} \\  
        $2^{20}$ & 213 MB & 28.73 / 25.37 & 0.9328 / 0.8404 & 0.0542 / 0.1677 \\
        \bottomrule
      \end{tabular}}
    }
\end{table}

\subsection{Evaluation}
We now evaluate various components of our algorithm.  Figure~\ref{fig:sampling} shows our adaptive sampling approach to training during the precomputation phase.  We first uniformly select samples on the scene based on the indirect light.  We then allocate additional samples in regions of high variance with respect to view, high-frequency regions, and those with high specular coefficients.  The total sample distribution is shown in the leftmost image.  


Table~\ref{tab:hashmapsize} shows an evaluation of different hashmap resolutions on the \textsc{Four Animals} scene.  We find empirically that a hash table size of $2^{19}$ produces the best results, and also has a reasonable storage size.  

In Table~\ref{tab:ablation_encodings}, we evaluated different encodings for the MLP.  They all performed similarly, but best PSNR was achieved with a spherical harmonic encoding only for $\omega_r$, with no impact on frame rate. Table~\ref{tab:ablation_endodings2} analyzes the encoding on just $\omega_r$ further on the more specular \textsc{Four Animals} by considering learned feature vectors to encode $\omega_r$ on the \textsc{Four Animals} scene. For these learned feature vector experiments, we converted $\omega_r$ to UV coordinates within a differentiable cubemap and learned a hash grid for each face, which we called DCE. To compare this with the SH embedding, we considered a high- and low-frequency hash grid configuration to explore the impact of potentially overfitting to training views. To further explore the contribution of encoding $\omega_r$, we also conducted experiments where the encoding is multiplied with the position and wavelet encodings in the initial CP decomposition phase to ablate its overall impact on the algorithm. In general, these approaches do not perform better than the simple spherical harmonic encoding, which we use for all of our results. A more thorough study on bandlimiting the angular component of these neural algorithms may be interesting as future work.

Figure~\ref{fig:histogramandspread} visualizes the wavelet statistics of a particular view. While the shape of the distribution we learn is different from the ground truth, most of the energy of these wavelets is nonetheless contained within relatively few entries. Note that this is for the full transport matrix; as in previous work~\cite{Ravi2}, for rendering an image we can make use of the most important wavelet coefficients in the lighting, dramatically reducing the number of wavelets needed to 64 in our case.

In Table~\ref{tab:ablation_hashgrid}, we compare the use of hashgrids for $S_m$ versus storing this information on vertices, using both learned feature vectors and spherical harmonics (maximum degree 8 using the traditional approach~\cite{Sloan1}), showing the benefit of using the volumetric hashgrid. We do this on the \textsc{Four Animals} scene, as this best showcases the challenges of storing the feature vectors on mesh vertices. We perform barycentric interpolation on the mesh vertices closest to the hit point. In the learned method, we apply a nonlinearity (softplus) to increase the representative capacity of the model. Spherical Harmonics perform worse than our method, lacking sharp reflection details and showing 
ringing and other artifacts on high-frequency light transport effects like caustics. This underscores that spherical harmonics are primarily a low-frequency representation. The learned feature vectors stored on the vertices perform better, but are noisier and still demonstrate inability to reconstruct caustics. Additionally, they require a well-subdivided mesh, so they would not be invariant to scaling scene complexity. Figure~\ref{fig:representation} shows a visual comparison.

Figure~\ref{fig:denoising} shows results before (middle) and after (right) denoising, indicating that our initial results are already high quality, but a small amount of wavelet noise can be removed by the standard OptiX denoiser.  The left image is a comparison to an equal time path-traced image without denoising which is substantially worse.  Note that denoising brute-force path tracing does not resolve complex interreflections, as shown in Fig~\ref{fig:optix}.

\begin{table}[t]
  \centering
   {\caption{\em \textbf{Ablation study of different encodings on the \textsc{Kitchen} scene.} OB refers to the one-blob encoding~\cite{mueller2019neural}; SH refers to maximum degree-4 spherical harmonics. While encoding everything with one-blob performs slightly better than spherical harmonics in some categories, we chose to use spherical harmonics as it was superior to encoding everything in PSNR (for best results after denoising) while being extremely close in the other metrics.}
   \label{tab:ablation_encodings}}
   \resizebox{\linewidth}{!}{
     {\begin{tabular}{l c c c} 
        \toprule
          Encoding & PSNR ($\uparrow$) & SSIM ($\uparrow$) & LPIPS ($\downarrow$) \\
        \midrule
        No encoding & 21.84 / 17.84 & 0.8667 / 0.1294 & 0.0871 / 0.7739 \\
        OB($n$), OB($\omega_r$), OB($\sigma$) & 35.75 / 33.35 & \textbf{0.9647} / \textbf{0.9179} & \textbf{0.0495} / \textbf{0.1365} \\
        Just OB($\omega_r$) & 35.74 / 33.38 & 0.9641 / 0.9166 & 0.0498 / 0.1402 \\
        Just SH($\omega_r$) & \textbf{35.83} / \textbf{33.42}  & \textbf{0.9647} / 0.9170 & 0.0498 / 0.1399 \\ 
        \bottomrule
      \end{tabular}}
    }
\end{table}

\begin{table}[t]
  \centering
   {\caption{\em \textbf{Ablation study of different encodings on solely $\omega_r$ on the \textsc{Four Animals} scene.} DCE refers to a differentiable hashgrid-based cubemap encoding the reflected direction; DCFE refers to a differentiable hashgrid-based cubemap encoding that is multiplied with the vertex and wavelet features (as a total factorization of the transport tensor).}
   \label{tab:ablation_endodings2}}
   \resizebox{\linewidth}{!}{
     {\begin{tabular}{l c c c} 
        \toprule
          Encoding & PSNR ($\uparrow$) & SSIM ($\uparrow$) & LPIPS ($\downarrow$) \\
        \midrule
        DCE($\omega_r$) (high freq.) & 28.79 / 25.65 & 0.9315 / 0.8536 & 0.0615 / 0.1766 \\
        DCFE($\omega_r$) (high freq.) & 17.71 / 13.31 & 0.8111 / 0.4442 & 0.1871 / 0.4454 \\
        DCE($\omega_r$) (low freq.) & 28.94 / 25.72 & 0.9324 / 0.8551 & 0.0591 / 0.1726 \\
        DCFE($\omega_r$) (low freq.) & 16.72 / 12.65 & 0.8089 / 0.4070 & 0.1811 / 0.4427 \\
        SH($\omega_r$) (Ours) & \textbf{29.62 / 26.21} & \textbf{0.9390 / 0.8642} & \textbf{0.0489 / 0.1609} \\
        \bottomrule
      \end{tabular}}
    }
\end{table}

\begin{table}[t]
  \centering
   {\caption{\em \textbf{Comparison of storing $S_m$ on mesh vertices versus a hashgrid.} We compare to a learned mesh vertex-based scheme and to spherical harmonics with maximum degree of 8~\cite{Sloan1}.
   }
   \label{tab:ablation_hashgrid}}
\resizebox{\linewidth}{!}{
  \begin{tabular}{l c c c} 
    \toprule
    Feature scheme & PSNR ($\uparrow$) & SSIM ($\uparrow$) & LPIPS ($\downarrow$) \\
    \midrule
    Max. Deg. 8 SH & 23.14 / 20.61 & 0.9147 / 0.8066 & 0.0655 / 0.2072 \\
    Mesh Vertices & 28.38 / 25.08 & 0.9282 / 0.8383 & 0.0675 / 0.2090 \\
    Hashgrid & \textbf{29.62 / 26.21} & \textbf{0.9390 / 0.8642} & \textbf{0.0489 / 0.1609} \\
    \bottomrule
  \end{tabular}
}
\end{table}

\begin{table}[t]
  \centering
   {\caption{\em \textbf{Quantitative comparison on PBRT Bathroom.} Our numbers after denoising are higher in all categories than NPRT, but our method finds difficulty in reconstructing perfect mirror effects. Results shown are full / indirect and evaluated on far views only. }
   \label{tab:highroughnessresults}}
   \resizebox{\linewidth}{!}{
     {\begin{tabular}{l c c c c} 
      \toprule
        Methods  & Denoising & PSNR($\uparrow$) & SSIM($\uparrow$) & LPIPS($\downarrow$) \\
        \midrule
        NPRT  & \xmark & 30.25 / 28.94 & 0.9335 / 0.8817 & 0.0904 / 0.1841 \\
        NPRT & \cmark & 30.44 / 29.32 & 0.9375 / 0.8968 & 0.0720 / 0.1254 \\
        Ours & \xmark & 30.75 / 29.72 & 0.9289 / 0.8839 & 0.1157 / 0.2056 \\
        Ours & \cmark & \textbf{31.14} / \textbf{30.24} & \textbf{0.9459} / \textbf{0.9116} & \textbf{0.0686} / \textbf{0.1129} \\
        \bottomrule
      \end{tabular}}
    }
\end{table}

\subsection{Limitations}
\begin{figure} [!!t]
  \centering
  \includegraphics[width=1.0\columnwidth]{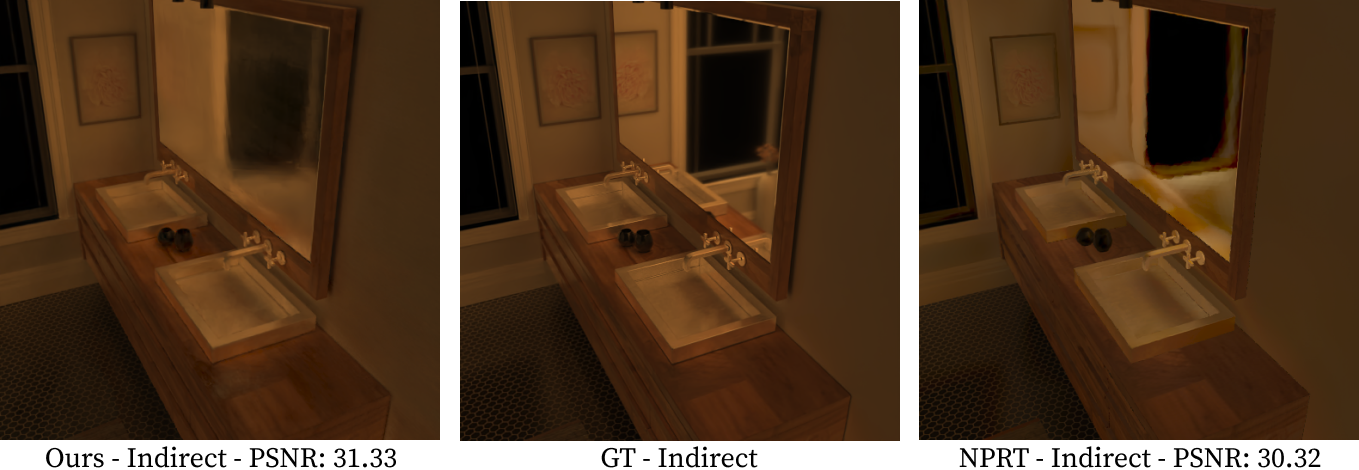}
  \caption{An example limitation of our method on PBRT Bathroom (numbers provided for image only; averages can be found in Table \ref{tab:highroughnessresults}). The mirror is Cook-Torrance with roughness 0.05. At a roughness this low both methods perform poorly, but while the overall highlights of NPRT are more inaccurate than ours, their capture of the reflection is more accurate (our method finds it hard to reconstruct the painting next to the mirror).}
  \label{fig:highroughnessfigure}
\vspace*{-.1in}
\end{figure}

Most limitations are inherited from previous PRT algorithms.  The results, while significantly higher quality than previous work are not perfect, since we use only 64 wavelets, and also approximate the transport coefficients.  Very high-frequency effects like mirrors are not perfectly reproduced (nor handled in previous techniques), and this can be seen in Figure \ref{fig:highroughnessfigure} where we evaluate our method on the PBRT Bathroom scene with the mirror set to 0.05 roughness. For reference, the full table of metrics is listed in Table \ref{tab:highroughnessresults}, evaluated on far views only. Some flicker can occasionally be seen in relighting as the selected wavelet coefficients change between environments (we minimize this by using area-weighted selection, which minimizes visual error by quickly resolving the diffuse colors as in \cite{Ravi2}).  Our optimization/training time for each scene can involve several hours, which is significantly higher than earlier non-learning approaches.  

Finally, our volumetric hashgrid, while significantly improving quality, does use more space than would a pure MLP or CNN approach, or in some cases a vertex-based method. 
Neural PRT does have a smaller model size/faster evaluation due to its weights being constrained within a single neural network (it uses only MLPs/CNNs rather than a feature grid).  
Our contribution is to provide substantially higher quality compared to Neural PRT, while using data sizes significantly lower than previous wavelet-based PRT methods -- our hashgrid is substantially more efficient than explicit transport matrix storage in early PRT work, often requiring at least an order of magnitude less storage.
An analogy can be made with NeRF-like models where the tradeoff is that feature fields can provide higher accuracy at the cost of higher required storage space (still much less than explicitly tabulated representations).  
An interesting future direction is to quantify the tradeoff between explicit feature fields and implicit methods in the PRT space.
\section{Conclusions and Future Work}
All-frequency relighting for indirect glossy reflections with changing
illumination and view has been one of the long-standing challenges for
precomputed radiance transfer, and real-time rendering in general.  In
this paper, we have taken an important step toward this goal, showing that a new approach leveraging modern MLP,
hashgrid, and novel factorization techniques can address the
challenge of glossy global illumination, obtaining the best of both traditional orthogonal Haar
wavelet decomposition and neural light transport approximation.  
In future work, we wish to consider alternative
factorizations and feature grids that may be more accurate and
compact, and alternatives to the hash-grid that can be computed
directly on the object/scene surface.  More broadly, this paper has
introduced a neural representation of 6D light transport that may be
applicable in many other areas including acquisition of the appearance
of real scenes, and for modeling of neural materials.  


\section{Acknowledgements}

We thank Peter-Pike Sloan, Alexandr Kuznetsov, Lingqi Yan, Ari Silvennoinen, Michał Iwanicki, Yash Belhe, Mohammad Shafiei, Pratul Srinivasan, Zhengqin Li and Alexander Mai for comments and discussions.
We additionally thank Alexander Mai, Falko Kuester, Mustafa Yaldız and Xiaoshuai Zhang for generously allowing us to use their compute resources, and we thank Gilles Rainer for answering questions.
This work was funded in part from NSF grants 2212085, 2100237 and 2120019, the Ronald L. Graham Chair and the UC San Diego Center for Visual Computing. We also acknowledge gifts from Google, Adobe, Qualcomm, Meta and a Sony Research Award.

\bibliographystyle{eg-alpha}
\bibliography{combined}

\end{document}